# Cardiotensor: A Python Library for Orientation Analysis and Tractography in 3D Cardiac Imaging

Joseph Brunet[1,2], Lisa Chestnutt[3], Matthieu Chourrout[1], Hector Dejea[2], Vaishnavi Sabarigirivasan[3,4], Peter D. Lee[1,5], Andrew C. Cook[3]

**Affiliations:**
1 Department of Mechanical Engineering, University College London, London, UK
2 European Synchrotron Radiation Facility, Grenoble, France
3 UCL Institute of Cardiovascular Science, London, UK
4 UCL Division of Medicine, University College London, London, UK
5 Research Complex at Harwell, Didcot, UK

## Summary

Understanding the architecture of the human heart requires analysis of its microstructural organization across scales. With the advent of high-resolution imaging techniques such as synchrotron-based tomography, it has become possible to visualize entire hearts at micron-scale resolution. However, translating these large, complex volumetric datasets into interpretable, quantitative descriptors of cardiac organization remains a major challenge. Here we present cardiotensor, an open-source Python package designed to quantify 3D cardiomyocyte orientation in whole- or partial-heart imaging datasets. It provides efficient, scalable implementations of structure tensor analysis, enabling extraction of directional metrics such as helical angle (HA), intrusion angle (IA), and fractional anisotropy (FA). The package supports datasets reaching teravoxel-scale and is optimized for high-performance computing environments, including parallel and chunk-based processing pipelines. In addition, cardiotensor includes tractography functionality to reconstruct continuous cardiomyocyte trajectories. This enables multi-scale myoaggregate visualization down to the myocyte level, depending on resolution. These capabilities enable detailed structural mapping of cardiac tissue, supporting the assessment of anatomical continuity and regional organization.

## Statement of Need

Despite major advances in high-resolution 3D imaging, there is a lack of open-source tools to analyze cardiomyocyte orientation in large volumetric datasets. Most established frameworks were developed for diffusion tensor MRI (DT-MRI), where orientation is inferred from local diffusion of water. Examples include MRtrix3 (Tournier et al., 2019), DIPY (Garyfallidis et al., 2014), and DSI Studio (Yeh, 2025). While powerful for diffusion-based neuro and cardiac applications (Mekkaoui et al., 2017), these packages are not designed to handle direct image-gradient–based orientation estimation or the teravoxel-scale datasets produced by synchrotron tomography, micro-CT, or 3D optical microscopy.

For non-diffusion imaging modalities, such as micro-CT (Reichardt et al., 2020), optical microscopy (Dileep et al., 2023; Garcia-Canadilla et al., 2022), and synchrotron tomography (Brunet et al., 2024; Dejea et al., 2019), researchers have historically relied on custom structure tensor implementations to estimate myoaggregate orientation directly from image intensity gradients. However, most of these are in-house codes, often unpublished or not scalable. Existing tools like OrientationJ (Fiji) and OrientationPy (Python) enable 2D and 3D structure tensor analysis (Navaee et al., 2023), but are not optimized for teravoxel-scale datasets, do not compute classical cardiac microstructure descriptors such as HA and IA, and do not support tractography for myoaggregate orientation mapping.

Cardiotensor is an open-source Python package specifically tailored to structure tensor analysis of large cardiac volumes. Rather than relying on diffusion modeling, it infers tissue orientation directly from image intensity gradients, making it applicable across a wide range of modalities and scales. Previous studies have demonstrated strong agreement between structure tensor–based orientation and DT-MRI–derived metrics when applied to the same human hearts (Teh et al., 2016). The package supports full pipelines from raw image stacks to myocyte orientation maps and tractography. Its architecture is optimized for large datasets, using chunked and parallel processing suitable for high-performance computing environments. The overall processing workflow, from input volumes to orientation maps and tractography, is summarized in Figure 1.

Cardiotensor has already been successfully applied in published work to characterize 3D cardiomyocyte architecture in healthy and diseased human hearts using synchrotron tomography (Brunet et al., 2024) on datasets over a terabyte in size. While conceived for cardiac imaging, the package is modality- and tissue-agnostic. Any volumetric dataset exhibiting coherent fibrous microstructure can be analyzed, including brain white matter, skeletal muscle, and tendon. This generality makes the library useful for both cardiovascular and broader anatomical or histological studies.

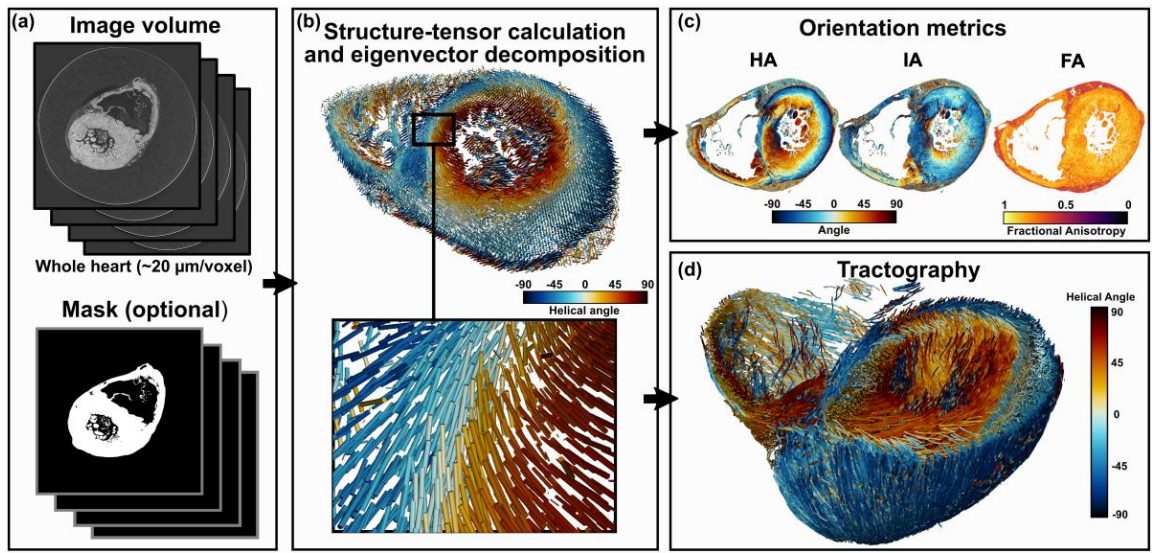

*Figure 1. Cardiotensor pipeline for 3D cardiac orientation analysis and tractography. (a) Input whole- or partial-heart volume with optional myocardial mask. (b) Local cardiomyocyte orientation estimated via 3D structure tensor and eigenvector decomposition. The third eigenvector field (smallest eigenvalue) is visualized as arrows color-coded by helical angle (HA); inset shows structure tensor orientation in the ventricular septum. (c) Transformation to a cylindrical coordinate system enables computation of voxel-wise HA, intrusion angle (IA), and fractional anisotropy (FA) maps. (d) Streamline tractography reconstructs continuous cardiomyocyte trajectories, color-coded by HA.*

## Implementation

Cardiotensor is implemented in Python and designed to efficiently process very large 3D cardiac imaging datasets. It relies primarily on NumPy (Van Der Walt et al., 2011) for numerical computation, with I/O accelerated by tifffile (Gohlke, 2020/2025), Glymur (Evans, 2013/2025), and OpenCV (Bradski, 2000). Dask (Rocklin, 2015) is used exclusively to parallelize file reading, while the core computations rely on Python's multiprocessing module for local parallelism. The package builds on the structure-tensor library (Jeppesen et al., 2021) to calculate the 3D structure tensor and perform eigenvector decomposition.

The package supports multiple use cases:

- Command-line workflows, which automate batch processing from a configuration file of terabyte-scale heart volumes and produce results as live plots or files saved to disk.
- Embedded use in larger Python analysis workflows, enabling flexible scripting and scalable execution on cluster environments.

Efficient computation is achieved through a chunk-based processing strategy with padding, which avoids edge artifacts. This architecture allows parallelization across computing clusters by splitting volumes into independent jobs, enabling cardiotensor to process whole-heart volumes in hours rather than days while maintaining practical memory requirements.

## Architecture

Cardiotensor is organized into five main modules, designed for clarity and scalability:

- **orientation**: Computes local cardiomyocyte (or other texture feature) orientation using a chunked 3D structure tensor pipeline, including eigenvalue decomposition, cylindrical coordinate rotation, and calculation of helical angle (HA), intrusion angle (IA), and fractional anisotropy (FA).
- **tractography**: Generates and filters streamlines tracing cardiomyocyte trajectories from the orientation field for myoaggregate-level reconstruction and analysis.
- **analysis**: Provides a GUI for regional quantification and plotting transmural profile.
- **visualization**: Supports interactive 3D visualization of vector fields and streamlines, HA color-coding, and export to VTK/ParaView for large-scale rendering.
- **utils**: Contains general utilities for I/O, image preprocessing, configuration parsing, and vector math, supporting the entire pipeline.

This modular architecture ensures reproducibility, maintainability, and easy integration into larger cardiac imaging workflows.

## Documentation and Usage

The documentation for cardiotensor is available online at:

https://josephbrunet.github.io/cardiotensor

The main components of the documentation are:

- Step-by-step walkthroughs for installation, first steps, and a guided example covering all available commands. A small example dataset and its corresponding mask are provided with the package.
- In-depth explanations of the core algorithms used in cardiotensor, including structure tensor theory, helical angle calculation, fractional anisotropy (FA), and tractography integration.
- Reference guides for the command-line interface, configuration file format, and public API.

## Acknowledgements

The authors would like to thank David Stansby for his guidance on the Python package structure, documentation framework, and best practices for scientific software development.

This work was supported in part by the Chan Zuckerberg Initiative DAF (grant 2022-316777), the Wellcome Trust (310796/Z/24/Z), and the Additional Ventures Single Ventricle Research Fund (grant 1019894).

The authors gratefully acknowledge ESRF beamtimes md1290 and md1389 on BM18 as sources of the data.

Peter D. Lee is a CIFAR MacMillan Fellow in the Multiscale Human program and acknowledges funding from a RAEng Chair in Emerging Technologies (CiET1819/10). This research is also based on work supported by a CIFAR Catalyst Award.

AC's research is enabled through the Noé Heart Centre Laboratories, which are gratefully supported by the Rachel Charitable Trust via Great Ormond Street Hospital Children's Charity (GOSH Charity). The Noé Heart Centre Laboratories are based in The Zayed Centre for Research into Rare Disease in Children, which was made possible thanks to Her Highness Sheikha Fatima bint Mubarak, wife of the late Sheikh Zayed bin Sultan Al Nahyan, founding father of the United Arab Emirates, as well as other generous funders.